\pdfoutput=1

\documentclass[twocolumn,english]{revtex4}
\usepackage{times,amssymb,amsmath,graphicx}
\usepackage[bookmarks=false,pdffitwindow=false,pdfstartview={FitH}]{hyperref}
\usepackage[T1]{fontenc}
\usepackage{babel}
\usepackage{color}

\begin{document}

%%  ======================================================================  %%
%%  TO JEST  >> W A Z <<  O DLUGOSCI 70+8 ZNAKOW I MA SIE MIESCIC W LINII   %%
%%  ======================================================================  %%

\title{
  Wiedemann-Franz law for massless Dirac fermions with implications
  for graphene
}

\author{Adam Rycerz}
\affiliation{Institute for Theoretical Physics,
  Jagiellonian University, \L{}ojasiewicza 11, PL--30348 Krak\'{o}w, Poland}

\date{April 16, 2021}

\begin{abstract}
In the 2016 experiment by Crossno {\em et al.\/}
[Science {\bf 351}, 1058 (2016)], 
electronic contribution to the thermal conductivity of graphene was found
to violate the well-known Wiedemann-Franz (WF) law for metals.
At liquid nitrogen temperatures, the thermal to electrical
conductivity ratio of charge-neutral samples was more than $10$ times higher
than predicted by the WF law, what was attributed to interactions between
particles leading to collective behavior described by hydrodynamics.
Here we show, by adapting the handbook derivation of the WF law
to the case of massless Dirac fermions, that significantly enhanced thermal
conductivity should appear also in few- or even sub-kelvin temperatures, where
the role of interactions can be neglected. The comparison with numerical
results obtained within the Landauer-B\"{u}ttiker formalism for rectangular
and disk-shaped (Corbino) devices in ballistic graphene is also provided.  
\end{abstract}

\maketitle

\section{Introduction}
Soon after the advent of graphene it become clear that this two-dimensional
form of carbon shows exceptional thermal conductivity, reaching the room
temperature value of $\sim{}5000\,$W$/$m$/$K \cite{Bal08}, being over
$10$ times higher than that of copper or silver \cite{dimfoo}.
Although the dominant contribution to the thermal conductivity originates
from lattice vibrations (phonons), particularly these corresponding to
out-of-plane deformations \cite{Alo13,Alo14} allowing graphene to outperform
more rigid carbon nanotubes, the electronic contribution  to the thermal
conductivity ($\kappa_{\rm el}$) was also found
to be surprisingly high \cite{Cro16} in relation to the electrical
conductivity ($\sigma$) close to the charge-neutrality point \cite{Kat12}. 
One can show theoretically that the electronic contribution dominates
the thermal transport at sub-kelvin temperatures \cite{Sus18}, but 
direct comparison with the experiment is currently missing.
Starting from a~few kelvins, up to the temperatures of about
$T\lesssim{}80\,$K, it is possible to control the temperatures of electrons
and lattice independently \cite{Cro16}, since the electron-phonon coupling
is weak, and to obtain the value of $\kappa_{\rm el}$ directly. 
Some progress towards extending the technique onto sub-kelvin temperatures
has been recently reported \cite{Dra19}. 

The Wiedemann-Franz (WF) law states that the ratio of $\kappa_{\rm el}$
to $\sigma$ is proportional to the absolute temperature \cite{Kit05}
\begin{equation}
  \label{wfmetal}
  \frac{\kappa_{\rm el}}{\sigma}={\cal L}T, 
\end{equation}
where the proportionality coefficient ${\cal L}$ is the Lorentz number. 
For ideal Fermi gas, we have
\begin{equation}
  \label{lornum0}
  {\cal L} = {\cal L}_0\equiv\frac{\pi^2}{3}\left(\frac{k_B}{e}\right)^2
  \simeq{} 
  2.443\times{}10^{-8}\ \mbox{W}\cdot\mbox{$\Omega$}\cdot\mbox{K}^{-2}.  
\end{equation}
For metals, Eq.\ (\ref{wfmetal}) with ${\cal L}\approx{}{\cal L}_0$
(\ref{lornum0}) holds
true as long as the energy of thermal excitations $k_B{}T\ll{}\varepsilon_F$,
with $\varepsilon_F$ being the Fermi energy.
What is more, in typical metals close to the room temperature
$\kappa_{\rm el}\gg{}\kappa_{\rm ph}$, with $\kappa_{\rm ph}$ being the phononic
contribution to the thermal conductivity, and even when approximating
the Lorentz
number as ${\cal L}\approx{}(\kappa_{\rm el}+\kappa_{\rm ph})/\sigma{}T$ one
restores the value of ${\cal L}_0$ (\ref{lornum0}) with a~few-percent accuracy. 

In graphene, the situation is far more complex, partly because $\kappa_{\rm el}
\ll{}\kappa_{\rm ph}$ (starting from few Kelvins) but mainly because unusual
properties of Dirac fermions in this system.
Experimental results of Ref.\ \cite{Cro16} show that the direct
determination of $\kappa_{\rm el}$ leads to 
${\cal L}/{\cal L}_0=10-20$ for $T=50-75\,$K near the charge-neutrality point.
Away from the charge-neutrality point, the value of
${\cal L}\approx{}{\cal L}_0$ is
gradually restored \cite{Kim16}.
Also, the Lorentz number is temperature-dependent, 
at a~fixed carrier density, indicating the violation of the WF law. 

High values of the Lorentz number (${\cal L}/{\cal L}_0>10$) were 
observed much earlier for semiconductors \cite{Gol56}, where the upper
limit is determined by the energy gap ($\Delta$) to temperature ratio,
${\cal L}_{\rm max}\approx{}(\Delta/2eT)^2$, 
but for zero-gap systems strong deviations from the WF law are rather
unexpected.
Notable exceptions are quasi one-dimensional Luttinger liquids, for which
${\cal L}/{\cal L}_0>10^4$ was observed \cite{Wak11}, and heavy-fermion
metals showing ${\cal L}<{\cal L}_0$ \cite{Tan07}. 

The peak in the Lorentz number appearing at the charge neutrality point
for relatively high temperatures (close to the nitrogen boiling point)
can be understood within a~hydrodynamic transport theory for graphene
\cite{Luc18,Zar19}.
However, it is worth to stress that for clean samples and much
lower temperatures, where the ballistic transport prevails,
one may still expect similar peaks
with the maxima reaching ${\cal L}_{\rm max}/{\cal L}_0\approx{}2-3$ and the
temperature-dependent widths. 

In this paper 
we show how to adapt the handbook derivation of the WF law \cite{Kit05}
in order to describe the violation of this law due to peculiar dispersion
relation and a~bipolar nature of graphene. The quantitative comparison with the 
Landauer-B\"{u}ttiker results is also presented, both for toy models of
the transmission-energy dependence, for which closed-form formulas for
${\cal L}$ are derived, and for the exact transmission
probabilities following from the mode-matching analysis for the rectangular
\cite{Kat06,Two06,Pra07} and for the disk-shaped \cite{Ryc09,Ryc10} samples.

The remaining part of the paper is organised as follows.
In Sec.\ \ref{wfgases} we recall the key points of the WF law derivation
for ideal Fermi gas, showing how to adapt them for massless fermions
in graphene.
In Sec.\ \ref{lanbutt}, the Landauer-B\"{u}ttiker formalism is introduced,
and the analytical results for simplified models for transmission-energy
dependence are presented. 
The Lorentz numbers for mesoscopic graphene systems, the rectangle and the
Corbino disk, are calculated in Sec.\ \ref{mesosys}. 
The conclusions are given in Sec.\ \ref{conclu}.

\section{Wiedemann-Franz law for ideal Fermi and Dirac gases}
\label{wfgases}

\subsection{Preliminaries}
\begin{figure}[!t]
  \includegraphics[width=0.45\linewidth]{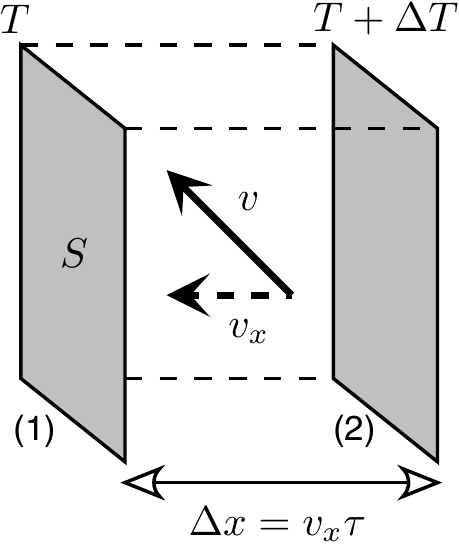}
  \caption{ \label{kapcvl:fig}
    Relation between the thermal conductivity ($\kappa$), heat capacity
    per unit volume ($C$), average particle velocity ($v$), and
    the mean-free path ($\ell$). The non-equilibrium heat flow occurs between
    the interfaces (1) and (2), with local temperatures $T$ and $T+\Delta{}T$,
    separated by a~distance $\Delta{}x=v_x{}\tau$ (with $v_x$ the mean velocity
    in $x$ direction and $\tau$ the relaxation time) and can be quantified by 
    $ \Delta{}Q = CS\Delta{}x\Delta{}T$.
    The corresponding  thermal conductivity is 
    $\kappa = \Delta{}Q\left(S\tau{}\Delta{}T/\Delta{x}\right)^{-1}
    = Cv_x^2\tau$.
    Substituting $v_x^2=v^2/d$, and $\ell = v\tau$, we obtain
    Eq.\ (\ref{kapcvl}) in the main text. 
  }
\end{figure}

The derivation of the WF law for metals \cite{Kit05} starts from
the relation between thermal conductivity of a~gas with its heat capacity per
unit volume ($C$) derived within kinetic theory of gases \cite{Kit05ch5},
which can be written as 
\begin{equation}
  \label{kapcvl}
  \kappa = \frac{1}{d}\,C{}v\,\ell, 
\end{equation}
where $d=1,2,3$ is the system dimensionality, $v$ is a~typical particle
velocity, and ${\ell}$ is the mean-free path (travelled between collisions
with boundaries or other particles). 
For the key points necessary to obtain Eq.\ (\ref{kapcvl}), 
see Fig.\ \ref{kapcvl:fig}. 
It is worth to notice  that the definition of $C$ in Eq.\ (\ref{kapcvl}),
used instead of a~familiar specific heat (per unit mass), allows to generalize
the reasoning onto the massless perticles easily. 

Next, the electrical conductivity in Eq.\ (\ref{wfmetal}) is expressed
via the Drude formula
\begin{equation}
  \label{sigdru}
  \sigma = \frac{n{}e^2\ell}{m_\star{}v},
\end{equation}
where $n=N/V$ is the carrier density (to be redefined later for a~bipolar system
containing electrons and holes), and $m_\star$ is the carrier effective mass.
We skip here the detailed derivation of Eq.\ (\ref{sigdru}), which can be
found in Ref.\ \cite{Kit05}; we only mention that it follows from Ohm's
law in the form ${\bf j}=\sigma{}{\bf E}$, with ${\bf j}$ the current
density and ${\bf E}$ the electric field, supposing that carriers of the
$\pm{}e$ charge and the $m_\star$ mass accelerate freely during the time
$\tau=\ell/v$ [with the symbols  $\ell$ and $v$ same as in
Eq.\ (\ref{kapcvl})]. This time, a~generalization for massless particles is
more cumbersome; we revisit this issue in Sec.\ \ref{wfdirgas}. 

The system volume, referred in definitions of $C$ and $n$, can be denoted
as $V=L^{d}$, with $L$ being linear dimension of a~box of gas.
In the SI units, the dimension of $C$ is J$/$(m$^d\cdot$K$)$, and the
unit of thermal conductivity is
\begin{equation}
  \label{unikap}
  [\,\kappa\,] = \frac{\mbox{W}}{\mbox{m}^{d-2}\cdot\mbox{K}}\, . 
\end{equation}
Similarly, the unit of electrical conductivity is 
\begin{equation}
  \label{unisig}
  [\,\sigma\,] = \frac{1}{\mbox{m}^{d-2}\cdot\mbox{$\Omega$}}\, . 
\end{equation}
In turn, the unit of length (m) vanishes in the $\kappa/\sigma$ ratio
occurring in Eq.\ (\ref{wfmetal}) and the WF law remains valid for arbitrary
$d$ (provided that the suppositions given explicitly in
Sec.\ \ref{wffergas} are satisfied.)
Unfortunately, in the literature on
graphene $\sigma$ is commonly specified in $\Omega^{-1}$ ($\equiv{}S$),
as follows from Eq.\ (\ref{unisig}) for $d=2$, but the values of $\kappa$
are reported in W$/$m$/$K, as for $d=3$ \cite{dimfoo}. Such an inconsistency
can be attributed to the fact that for the thermal conductivity of multilayer
graphenes linear scaling with the number of layers remains a reasonable 
approximation \cite{Alo14apl}, yet the behavior of electrical conductivity
is far more complex \cite{Kos10,Nam17} even for bilayers \cite{Sus20b}.

\subsection{The Fermi gas in metals} 
\label{wffergas}
The calculation of  $C$ in Eq.\ (\ref{kapcvl}) employs the free
Fermi gas approximation for electrons in a~metal.
In this approximation, one assumes that leading contributions
to thermodynamic properties originate from a~thin layer around the
Fermi surface. For instance, a~contribution to the internal energy can
be written as
\begin{align}
  \label{deluel}
  \Delta{}U_{\rm el} &= \int_{\varepsilon_F-\Lambda}^{\varepsilon_F+\Lambda}
  d\varepsilon{}D(\varepsilon)f(\varepsilon)\varepsilon \nonumber \\
  &\approx \text{const.} +
  2D(\varepsilon_F)\left(k_B{}T\right)^2
  \int_{0}^{\infty}dx\frac{x}{e^x+1},
\end{align}
where $\varepsilon_F$ is the Fermi energy, $2\Lambda$ is
the relevant energy interval considered ($\varepsilon_F\gg{}\Lambda
\gg{}k_B{T}$), $D(\varepsilon)$ is the density of states per unit volume
(i.e., the number of energy levels lying in the interval of $\varepsilon,
\dots,\varepsilon+d\varepsilon$ is  $VD(\varepsilon)d\varepsilon\,$),
and $f(\varepsilon)$ is the Fermi-Dirac distribution function 
\begin{equation}
  \label{ferdir}
  f(\varepsilon)=\frac{1}{e^{{(\varepsilon-\mu)}/{k_BT}}+1}. 
\end{equation}

In a~general case, the chemical potential in Eq.\ (\ref{ferdir}) is adjusted
such that the particle density
\begin{equation}
  \label{nmuint}
  n(\mu) = \int_0^\infty{}d\varepsilon{}D(\varepsilon)f(\varepsilon)
\end{equation}
take a~desired value $n(\mu)\equiv{}n$, defining the temperature-dependent
chemical potential $\mu=\mu(T)$. Here, the constant-density of states
approximation, $D(\varepsilon)\approx{}D(\varepsilon_F)$ for
$\epsilon_F-\Lambda\leqslant{}\epsilon\leqslant\epsilon_F+\Lambda$
imposed in the rightmost expression in Eq.\ (\ref{deluel}), is equivalent
to $\mu\approx\varepsilon_F$ \cite{sommexp}. 

Definite integral in Eq.\ (\ref{deluel}) is equal to
\begin{equation}
  \label{zeta3}
  \int_{0}^{\infty}dx\frac{x}{e^x+1} = \frac{1}{2}\zeta(2) = \frac{\pi^2}{12},
\end{equation}
where the Riemann zeta function
\begin{equation}
  \zeta(z) = \sum_{p=1}^{\infty}\frac{1}{p^{-z}}, \ \ \ \ \ \ 
  \mbox{Re}\,z>1, 
\end{equation} 
is introduced to be used in forthcoming expressions. 

Differentiating  $\Delta{}U_{\rm el}$ (\ref{deluel}) over temperature, one
gets approximating expression for the electronic heat capacity
\begin{equation}
  \label{celfer}
  C_{\rm el}\approx{}\frac{\pi^2}{3}D(\varepsilon_F)k_B^2{}T. 
\end{equation}
In fact, the factor of $\pi^2/3$ in Eq.\ (\ref{celfer}) is the same as
appearing in the Lorentz number ${\cal L}_0$ (\ref{lornum0}),
what is shown in a~few remaining steps below. 

For an isotropic system with parabolic dispersion relation
\begin{equation}
  \label{paradisp}
  \varepsilon_{\bf k} = \frac{\hbar^2{}k^2}{2m_{\star}}, 
\end{equation}
bounded in a~box of the volume $V=L^d$ with periodic boundary conditions,
the wavevector components  ${\bf k}=(k_i)$ take discrete values of 
$k_i=0,\pm\frac{2\pi}{L},\pm\frac{4\pi}{L},\dots$ (with $i=x,y,z$ for $d=3$).
Calculation of the density of states in $d=1$, $2$, $3$ dimensions
is presented in numerous handbooks \cite{Zeg11}; here, we use a~compact
form referring to the particle density on the Fermi level
\begin{equation}
  \label{defnef}
  D(\varepsilon_F) = \frac{d}{2}\frac{n(\varepsilon_F)}{\varepsilon_F},
\end{equation} 
where $n(\varepsilon_F)=\int_0^{\varepsilon_F}D(\varepsilon)d\varepsilon$ 
representing the $T\rightarrow{}0$ limit of Eq.\ (\ref{nmuint}).  
Substituting $D(\varepsilon_F)$, given by Eq.\ (\ref{defnef}), into
Eq.\ (\ref{celfer}) we obtain
\begin{equation}
  C_{\rm el}\approx{}\frac{\pi^2{}d}{6}\,\frac{nk_B^2{}T}{\varepsilon_F}. 
\end{equation}

Now, taking $\epsilon_F=\frac{1}{2}m_\star{}v_F^2$ with the Fermi velocity
\begin{equation}
  \label{vfnrel}
  v_F = \frac{1}{\hbar}\left.
  \frac{\partial{}\varepsilon_{\bf k}}{\partial{}k}
  \right|_{k=k_F}
  =\frac{\hbar{}k_F}{m_\star}, 
\end{equation}
and the Fermi wavevector $k_F=\sqrt{2m_\star{}\epsilon_F/\hbar^2}$,
we further set $v=v_F$ in Eq.\ (\ref{kapcvl}), obtaining
\begin{equation}
  \label{kapfer}
  \kappa_{\rm el} \approx \frac{\pi^2}{3}\,\frac{nk_B^2{}T}{m_\star{}v_F}\ell. 
\end{equation}
It is now sufficient to divide Eqs.\  (\ref{kapfer}) and (\ref{sigdru})
side-by-side to derive the WF law as given by Eqs.\ (\ref{wfmetal}) and
(\ref{lornum0}).

As mentioned earlier, the result for free Fermi gas is same for arbitrary
dimensionality $d$. More careful analysis also shows that the parabolic
dispersion of Eq.\ (\ref{paradisp}) is not crucial, provided that the Fermi
surface is well-defined, with an (approximately) constant $D(\varepsilon)>0$
in the vicinity of $|\varepsilon-\varepsilon_F|\lesssim{}k_B{}T$, and that
the effective mass $0<m_\star<+\infty$.
In the framework of Landau's Fermi-liquid (FL) theory, the reasoning
can be extended onto effective quasiparticles, 
and the validity of the WF law is often considered as a~hallmark of the FL
behavior \cite{Mah13,Lav19}. 

The suppositions listed above are clearly not satisfied
in graphene close to the charge-neutrality point.

\subsection{The Dirac gas in graphene} 
\label{wfdirgas}

The relation between thermal conductivity and heat capacity given
by Eq.\ (\ref{kapcvl}) holds true for both massive and massless particles.
A~separate issue concerns the Drude formula (\ref{sigdru}), directly
referring to the effective mass, an adaptation of which for massless Dirac
fermions requires some attention.

The Landauer-B\"{u}ttiker conductivity of ballistic graphene, first
calculated analytically employing a~basic mode-matching technique
\cite{Kat06,Two06,Pra07} and then confirmed in several experiments
\cite{Mia07,Dan08}, is given solely by fundamental constants 
\begin{equation}
  \label{sig0gra}
  \sigma_0=\frac{4e^2}{\pi{}h}. 
\end{equation}
Remarkably, for charge-neutral graphene both the carrier concentration
and the effective mass vanish; a~finite (and nonzero) value of
$\sigma_0$ (\ref{sig0gra}) may therefore
be in accord with the Drude formula, at least in principle.  

In order to understand the above conjecture, we refer to the approximate
dispersion relation for charge carriers in graphene, showing up so-called
Dirac cones,
\begin{equation}
  \label{condir}
  E = \pm{}\hbar{}v_Fk. 
\end{equation}
The value of the Fermi velocity $v_F\approx{}10^6\,$m/s is now
energy-independent, being determined by the nearest-neighbor hopping
integral on a~honeycomb lattice ($t_0=2.7\,$eV) and the lattice constant
($a=0.246\,$nm) via 
\begin{equation}
  \label{vft0a}
  \hbar{}v_F=\frac{\sqrt{3}}{2}t_0{}a. 
\end{equation}

Charge carriers in graphene are characterized by an additional (next
to spin) quantum number, the so-called valley index. This leads to an
additional twofold degeneracy of energy levels, which needs to taken
into account when calculating the density of states, 
\begin{equation}
  \label{depsgra}
  D(\varepsilon) = \frac{2|\varepsilon|}{\pi{}(\hbar{}v_F)^2}. 
\end{equation}
Subsequently, the carrier concentration at $T=0$ is related to the Fermi
energy (and the Fermi wavevector) via
\begin{equation}
  \label{n0gra}
  n=\int_0^\varepsilon{}D(\varepsilon')d\varepsilon' = 
  \frac{\varepsilon^2}{\pi{}(\hbar{}v_F)^2} = \frac{k^2}{\pi}. 
\end{equation}
In the above we intentionally omitted the $F$ index for symbols denoting
the Fermi energy and the Fermi wavevector to emphasize that they can be 
tuned (together with the concentration) by electrostatic gates, while
the Fermi velocity $v_F$ (\ref{vft0a}) is a~material constant
\cite{strafoo}. 

Despite the unusual dispersion relation, given by Eq.\ (\ref{condir}),
the relevant effective mass describing the carrier dynamics
in graphene is the familiar cyclotronic mass
\begin{equation}
  \label{mcygra}
  m_C = \frac{\hbar^2}{2\pi}
  \frac{\partial{\cal A}(\varepsilon)}{\partial\varepsilon} =
  \frac{\hbar{}k}{v_F}, 
\end{equation}
where ${\cal A}(\varepsilon)$ denotes the area in momentum space $(k_x,k_y)$
bounded by the equienergy surface for a~given Fermi energy ($\varepsilon$).
It is easy to see that for two-dimensional system, with fourfold degeneracy
of states, we have ${\partial{\cal A}(\varepsilon)}/{\partial\varepsilon}
=\pi^2{}D(\varepsilon)$; substituting $D(\varepsilon)$ given by
Eq.\ (\ref{depsgra}) leads to the rightmost equality in Eq.\ (\ref{mcygra}).
Remarkably, the final result is formally identical with the rightmost
equality in Eq.\ (\ref{vfnrel}) for free Fermi gas
(albeit now the effective mass, but not the Fermi velocity, depends
on the Fermi energy). 
%albeit the meaning of symbols is different than for the Dirac gas. 

Assuming the above carrier density $n$ (\ref{n0gra}), and the effective
mass $m_\star$ (\ref{mcygra}), and comparing the universal conductivity
$\sigma_0$ (\ref{sig0gra}) with the Drude formula (\ref{sigdru}), we
immediately arrive to the conclusion that mean-free path for charge
carriers in graphene is also energy-dependent, taking the asymptotic form
\begin{equation}
  \label{elleff}
  \ell_{\rm eff}(\varepsilon) \simeq 
  \frac{2}{\pi{}k}=\frac{2\hbar{}v_F}{\pi{}\varepsilon},
  \qquad \text{for }\ \ \varepsilon\rightarrow{}0. 
\end{equation}
Strictly speaking, in the $\varepsilon\rightarrow{}0$ limit is have
$n\rightarrow{}0$, i.e., no free charge carriers, and the transport 
is governed by evanescent waves \cite{Kat12}.
The universal value of $\sigma_0$ (\ref{sig0gra}) indicates a~peculiar
version of the tunneling effect appearing in graphene, in which
the wavefunction shows a~power-law rather then exponential decay
with the distance \cite{Ryc09}, resulting in the enhanced charge (or energy)
transport characteristics. 
Therefore, the mean-free path should be regarded as an effective quantity,
allowing one to reproduce the measurable characteristics in the
$\varepsilon\rightarrow{}0$ limit. 
Away from the charge-neutrality point, i.e., for $|\varepsilon|\gg{}\pi\hbar{}
v_F/L$ (with the geometric energy quantization $\sim\pi\hbar{}v_F/L$),
graphene behaves as a~typical ballistic conductor, with $\ell_{\rm eff}\sim{}L$.
We revisit this issue in Sec.\ \ref{mesosys},
where the analysis starts from actual $\sigma(\varepsilon)$ functions for
selected mesoscopic systems, but now the approximation given by 
Eq.\ (\ref{elleff}) is considered as a~first. 

We further notice that the form of $\ell_{\rm eff}(\varepsilon)$ in Eq.\
(\ref{elleff}) is formally equivalent to the assumption of linear
relaxation time on energy dependence in the Boltzmann equation, 
proposed by Yoshino and Murata \cite{Yos15}. 

In the remaining part of this section, we derive explicit forms of the
thermal conductivity $\kappa$ and the Lorentz number ${\cal L}$,
pointing out the key differences appearing in comparison to
the free Fermi gas case (see Sec.\ \ref{wffergas}). 

The calculations are particularly simple for charge-neutral graphene 
($n=\varepsilon=0$), which is presented first.
Although we still can put $v=v_F$ in Eq.\ (\ref{kapcvl}), since the
Fermi velocity is energy-independent, the constant-density of states
approximation applied in Eq.\ (\ref{deluel}) in now invalid.
(Also, for $T>0$ we cannot put $\varepsilon_F\gg{}k_B{}T$ now.)
In turn, the expression for heat capacity $C$ needs to be re-derived.

For charge-neutral graphene at $T>0$, contributions from thermally excited
electron and holes are identical, it is therefore sufficient to calculate
the former 
\begin{multline} 
  \label{uegra}
  U_e(T) = \int_0^{\infty}d\varepsilon{}
  D(\varepsilon)f(\varepsilon)\,\varepsilon \\
  = 
  \frac{2(k_B{}T)^3}{\pi{}(\hbar{}v_F)^2}
  \int_0^{\infty}dx\,\frac{x^2}{e^x+1}. 
\end{multline}
Again, the integral in the rightmost expression in Eq.\ (\ref{uegra}) can
be expressed via the Riemann zeta function, and is equal to 
\begin{equation}
  \int_0^{\infty}dx\frac{x^2}{e^x+1} = \frac{3}{2}\zeta(3) \approx
  1.8031. 
\end{equation}
Differentiating Eq.\ (\ref{uegra}) with respect to $T$, and multiplying
by a~factor of $2$ due to the contribution from holes in the valence band, 
we obtain the heat capacity 
\begin{equation}
  \label{capgra}
  C = \frac{18\,\zeta(3)}{\pi}\,\frac{k_B^3{}T^2}{(\hbar{}v_F)^2}. 
\end{equation}

It remains now to calculate the effective mean-free path $\ell$ to be
substituted to Eq.\ (\ref{kapcvl}). We use here the asymptotic form of 
$\ell_{\rm eff}(\varepsilon)$ (\ref{elleff}), replacing the $\varepsilon^{-1}$
factor by its overage over the grand canonical ensemble, namely
\begin{equation}
  \langle\epsilon^{-1}\rangle_{T>0} = 
  \frac{\int_0^{\infty}d\epsilon{}D(\epsilon)f(\epsilon)\epsilon^{-1}}{\int_0^{\infty}d\epsilon{}D(\epsilon)f(\epsilon)} = 
  \frac{12\ln{}2}{\pi^2{}k_B{}T}. 
\end{equation}
Substituting the above, together with the heat capacity $C$ (\ref{capgra})
into Eq.\ (\ref{kapcvl}), we get
\begin{equation}
  \kappa = \frac{432\ln{}2\,\zeta(3)}{\pi^3{}h}k_B^2{}T,
\end{equation}
and 
\begin{equation}
  \label{wfgra}
  \frac{\kappa}{\sigma_0{}T} = \frac{108\ln{}2\,\zeta(3)}{\pi^2}\left(\frac{k_B}{e}\right)^2{}
  \approx 2.7714\times{}{\cal L}_0, 
\end{equation}
with ${\cal L}_0$ being the Fermi-gas result given by Eq.\ (\ref{lornum0}). 

A~simple reasoning, presented above, indicates that the $\kappa/\sigma$
ratio is significantly enhanced in charge-neutral graphene, comparing
to the free Fermi gas.
However, the WF law is still satisfied, since the Lorentz number
given by Eq.\ (\ref{wfgra}) is temperature-independent.
The situation becomes remarkably different for graphene away from
the charge-neutrality point, which is studied next.

\begin{figure}[!t]
  \includegraphics[width=\linewidth]{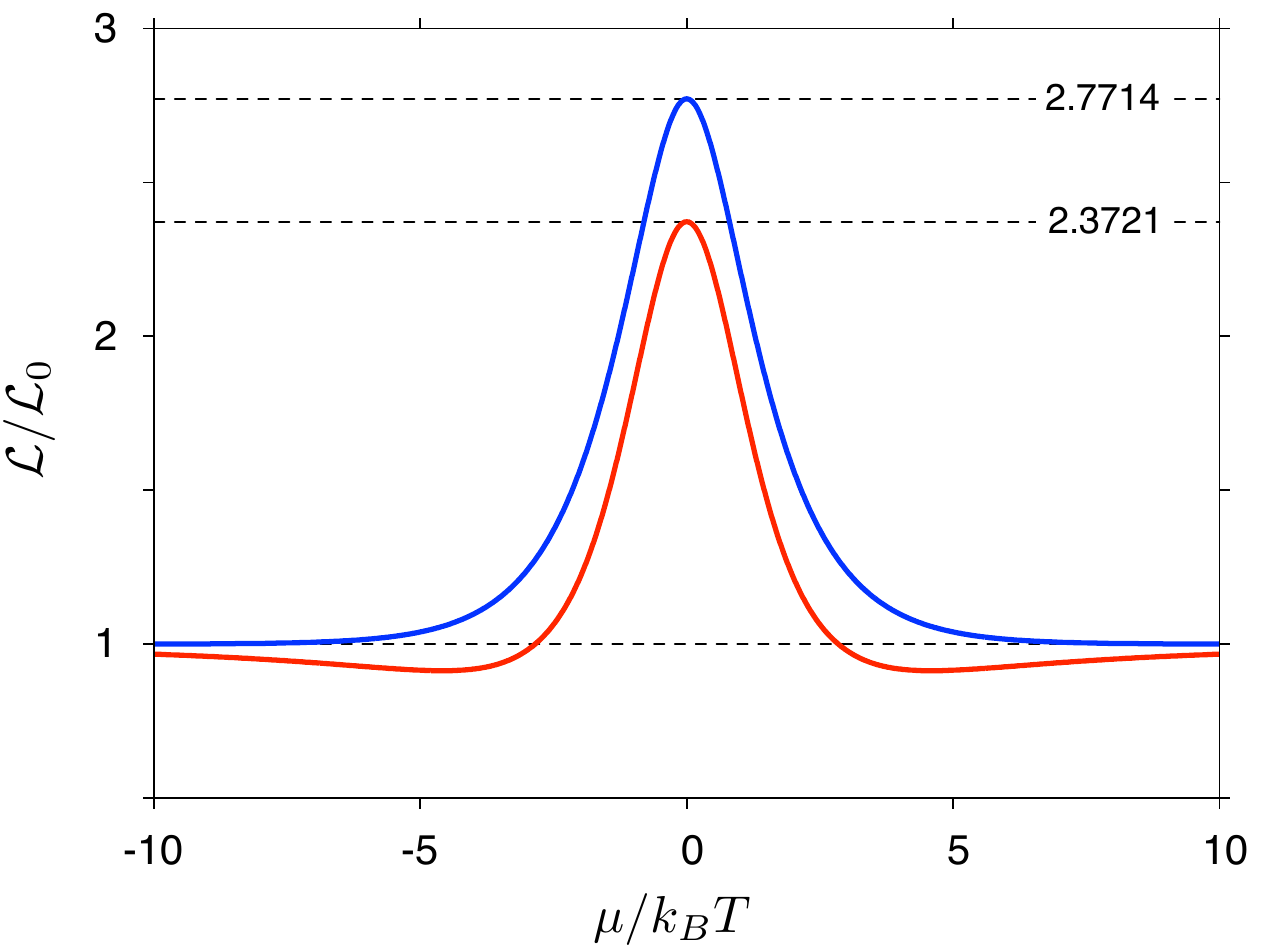}
  \caption{ \label{wfdirac}
    The Lorentz number ${\cal L}=\kappa_{\rm el}/(\sigma{}T)$ for massless
    Dirac fermions as a~function of the chemical potential.
    Solid lines represent the approximations given by Eq.\ (\ref{lorgra1})
    [blue line] and Eq.\ (\ref{lorevials}) [red line].
    Dashed lines (top to bottom) depict the two
    corresponding $\mu=0$ values, and the value of
    ${\cal L}_0=(\pi^2/3)\,k_B^2/e^2$ representing the Wiedemann-Franz law
    restored in the $|\mu|\gg{}k_BT$ limit. 
  }
\end{figure}

Without loss of generality, we suppose $\mu>0$ (the particle hole-symmetry
guarantees that measurable quantities are invariant upon
$\mu\rightarrow{}-\mu$). 
The internal energy $U(T)$ now consists of contributions from majority
carries (electrons), with $\varepsilon>\mu$, and minority carriers (holes), 
with $\varepsilon<\mu$,
\begin{align}
  U(T) &= U_e+U_h
  \nonumber \\
  &= \int_{\mu}^{\infty}d\varepsilon\,
  D(\varepsilon)
  \frac{\varepsilon-\mu}{\exp\left[(\varepsilon-\mu)/k_B{}T\right]+1}
  \nonumber \\
  &+ \int_{-\infty}^{\mu}d\varepsilon\,
  D(\varepsilon)
  \frac{\mu-\varepsilon}{\exp\left[(\mu-\varepsilon)/k_B{}T\right]+1},  
\end{align}
where $D(\varepsilon)$ is given by Eq.\ (\ref{depsgra}). 
The heat capacity can be written as
\begin{align}
  C &= \frac{\partial{}U}{\partial{}T}
  = \frac{1}{4k_B{}T^2}\int_{-\infty}^{\infty}d\varepsilon\,
  D(\varepsilon)\frac{(\varepsilon-\mu)^2}{{\cosh}^2\left[ %
  (\varepsilon-\mu)/k_B{}T\right]}
  \nonumber \\
  &= \frac{2k_B^3{}T^2}{\pi{}(\hbar{}v_F)^2}\,F(y), 
\end{align}
where we have defined 
\begin{align}
\label{fydef}
  F(y) &= \int_y^{\infty}\frac{dx\,x^3}{\cosh{}x+1}
  + y\int_0^y\frac{dx\,x^2}{\cosh{}x+1}
  \nonumber \\ 
  &= \frac{\pi^2}{3}y - y^3 + y^2\ln\left(2\cosh{}y+2\right)
  \nonumber \\ 
  &- 8\,\mbox{Li}_2(-e^{-y}) - 12\,\mbox{Li}_3(-e^{-y}), 
\end{align}
with $y=\mu/k_B{}T>0$ and Li$_s(z)$ being the polylogarithm function
\cite{Old09}. 

Similarly, the mean-free path can be calculated as
\begin{align}
  \langle\ell_{\rm eff}\rangle &=
  \frac{2\hbar{}v_F}{\pi}\left\langle|\epsilon|^{-1}\right\rangle_{\mu>0,T>0}
  \nonumber \\
  &= \frac{2\hbar{}v_F}{\pi{}k_B{}T}\,G(y), 
\end{align}
where
\begin{align}
\label{gydef}
  G(y) &= \ln{}2\times\left(\, \int_y^{\infty}dx\frac{x}{e^x+1}
  +y\int_0^y\frac{dx}{e^x+1} \,\right)^{-1}
  \nonumber \\
  &= \ln{}2\times\bigg[\,
    y^2+y\ln\left(e^{-y}+1\right) - \mbox{Li}_2\left(-e^{-y}\right)
    \nonumber \\
    &\qquad\qquad  {} -y\ln\left(e^{y}+1\right) + y\ln{}2
  \,\bigg]^{-1}, 
\end{align}
and $y=\mu/k_B{}T$ again. 

Hence, the Lorentz number for $\mu>0$ is given by
\begin{equation}
  \label{lorgra1}
  {\cal L} = \frac{\kappa}{\sigma_0{}T} =
  F(y)\,G(y) \left(\frac{k_B}{e}\right)^2,  
\end{equation}
with $F(y)$ and $G(y)$ given by Eqs.\ (\ref{fydef}) and (\ref{gydef}).
The Lorentz number given by Eq.\ (\ref{lorgra1}) is depicted
in Fig.\ \ref{wfdirac}. 
It is straightforward to show that in the $y\rightarrow{}0$ limit one 
obtains the value given by Eq.\ (\ref{wfgra}) for $\mu=0$;
also, for $y\rightarrow{}\infty$ we have ${\cal L}\rightarrow{}{\cal L}_0$,
restoring a~standard form of the WF law for metals. 
However, for $0<y<+\infty$, a~fixed value of $\mu$ (or $n$) corresponds
to $y$ (and thus ${\cal L}$) varying with temperature; namely, 
the violation of the WL law occurs.

\section{Landauer B\"{u}ttiker formalism and simplified models}
\label{lanbutt}

\subsection{The formalism essential}
In the Landauer-B\"uttiker description transport properties of a~mesoscopic
system, attached to the leads, are derived from the transmission-energy
dependence ${\cal T}(\varepsilon)$, to be found by solving the scattering
problem \cite{Lan57,But85,But86,But88}.
In particular, the Lorentz number can be written as \cite{Esf06}
\begin{equation}
  \label{lornano}
  {\cal L} = \frac{\kappa_{\rm el}}{\sigma{}T} =
  \frac{L_0L_2-L_1^2}{e^2{}T^2{}L_0^2},
\end{equation}
where $L_n$ (with $n=0,1,2$) are given by
\begin{equation}
  \label{llndef}
  L_n = \frac{g_sg_v}{h}\int{}d\varepsilon\,{\cal T}(\varepsilon)
  \left(-\frac{\partial{}f}{\partial{}\varepsilon}\right)
  (\varepsilon-\mu)^n,
\end{equation}
with $g_s=g_v=2$ denoting spin and valley degeneracies in graphene,
and the Fermi-Dirac distribution function $f(\varepsilon)$ given
by Eq.\ (\ref{ferdir}).
It is easy to show that energy-independent transmission
($\,{\cal T}(\varepsilon)=\ $const$\,$)
leads to ${\cal L}={\cal L}_0$ (\ref{lornum0}).

\subsection{Simplified models}
Before calculating ${\cal T}(\varepsilon)$ directly for selected systems
in Sec.\ \ref{mesosys}, we first discuss basic consequences of some model
${\cal T}(\varepsilon)$ functions for ${\cal L}$.

For instance, the linear transmission-energy dependence (i.e.,
$\,{\cal T}(\varepsilon) \propto{}|\varepsilon|\,$), allows one to obtain 
a~relatively short formula for ${\cal L}$ at arbitrary doping \cite{Sus18},
namely 
\begin{align}
\label{lorevials}
  {\cal L} &= \Bigg\{\,
  \frac{ {\pi^2}y+y^3-12\,\mbox{Li}_3(-e^{-y})}{\ln\left(2\cosh{y}+2\right)}
  \nonumber \\
  & \qquad - \left[\,
  \frac{{\pi^2}/{3}+y^2+4\,\mbox{Li}_2(-e^{-y})}{\ln\left(2\cosh{y}+2\right)}
  \,\right]^2  
  \Bigg\}\left(\frac{k_B}{e}\right)^2,   
\end{align}
with $y=\mu/k_B{}T$. For $y=0$, the Lorentz number given by
Eq.\ (\ref{lorevials}) takes the value of 
\begin{equation}
  \label{lor0lin}
  {\cal L}(0) = 
  \frac{9\,\zeta(3)}{2\ln{}2}\left(\frac{k_B}{e}\right)^2 
  \approx 2.3721\times{}{\cal L}_0,  
\end{equation}
being close to that given in Eq.\ (\ref{wfgra}). The approximation
given in Eq.\ (\ref{lor0lin}) was earlier put forward in the context
of high-temperature supeconductors also showing the linear
transmission-energy dependence \cite{Sha03}. 

Numerical values of ${\cal L}(y)$ are presented in Fig.\ \ref{wfdirac}. 
Remarkably, ${\cal L}(y)$ obtained from Eq.\ (\ref{lorgra1}) [blue line]
is typically $20-30\%$ higher than obtained Eq.\ (\ref{lorevials}) [red line]. 
The deviations are stronger near $|\mu|/k_B{}T\approx{}4.5$, where the latter
shows broad minima absent for the former. Above this value,
${\cal L}(y)$ obtained from Eq.\ (\ref{lorgra1}) approaches ${\cal L}_0$
from the top, whereas ${\cal L}(y)$ obtained from 
Eq.\ (\ref{lorevials}) approaches ${\cal L}_0$ from the bottom.
Also, the right-hand side of Eq.\ (\ref{lorgra1}) converges much faster
to ${\cal L}_0$ for $|\mu|\gg{}k_B{}T$ than the right-hand side of 
Eq.\ (\ref{lorevials}). 

In both cases, the Lorentz number enhancement at the charge-neutrality
point ($\mu=0$) is significant,
and the violations of the WF law for $\mu\neq{}0$ is apparent.
A~relatively good agreement between the two formulas is striking:
Although both derivations have utilized the linear dispersion of the Dirac
cones, being link to $D(\varepsilon)$ given by Eq.\ (\ref{depsgra})
in the first case,
or to the $T(\varepsilon)\propto{}|\varepsilon|$ assumption in the
second case (see Sec.\ \ref{mesosys} for further explanation), 
but only the derivation of Eq.\ (\ref{lorgra1}) incorporates
the information about the universal conductivity ($\sigma=\sigma_0$).
We can therefore argue that the ${\cal L}$ enhancement occurs in graphene
due to the linear dispersion rather then due the transport via evanescent
waves (being responsible for $\sigma=\sigma_0$ at $\mu=0$).

\begin{figure}[!t]
  \includegraphics[width=\linewidth]{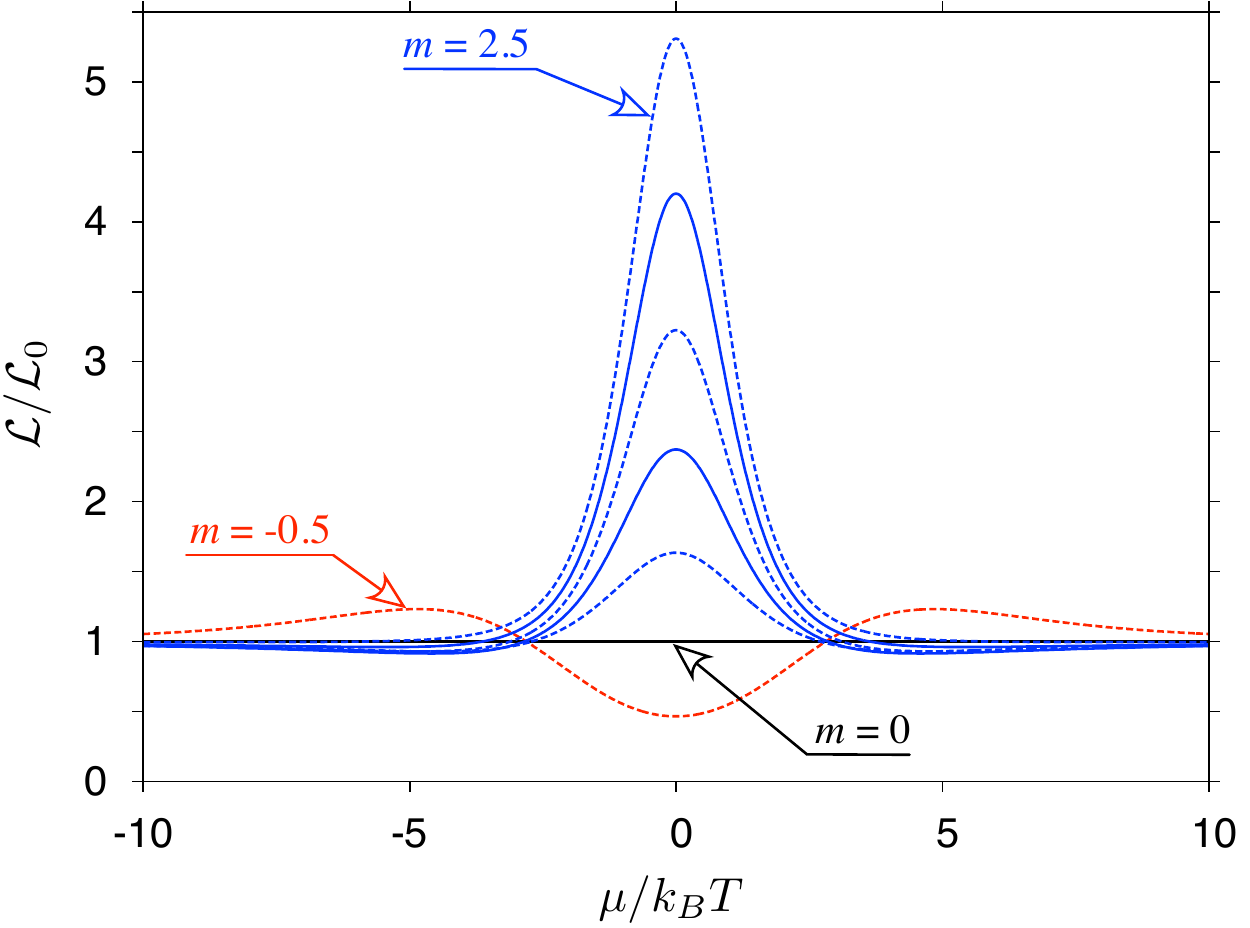}
  \caption{ \label{wftoys}
    The Lorentz number for model transmission-energy dependence
    ${\cal T}(\varepsilon)$ given by Eq.\ (\ref{temodm}) with $m$ varied
    from $-0.5$ to $2.5$ with the steps of $0.5$ displayed as a~function
    of the chemical potential. Solid (dashes) lines mark integer
    (non-integer) $m$. 
  }
\end{figure}

\begin{figure}[!t]
  \includegraphics[width=\linewidth]{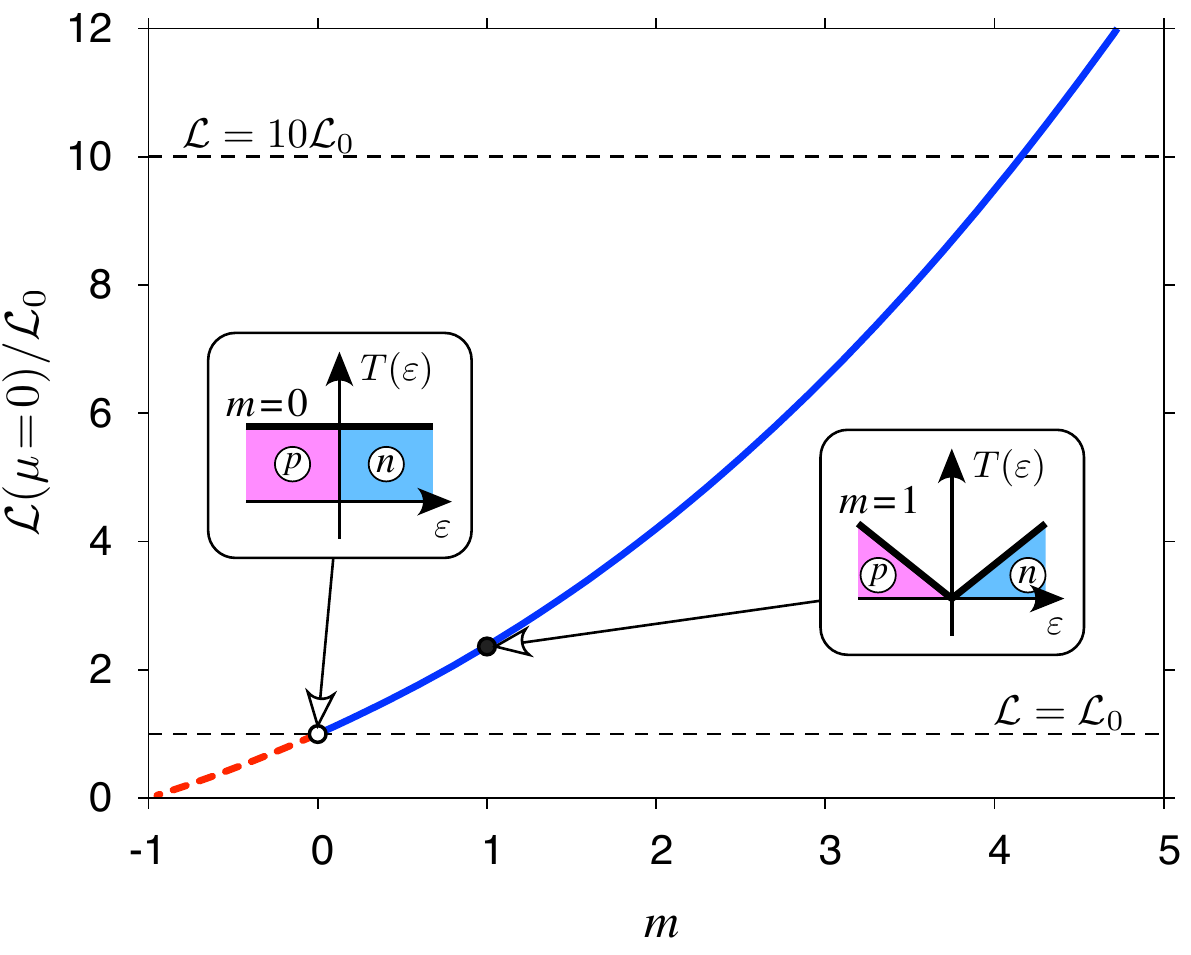}
  \caption{ \label{lormax}
    Maximal (solid blue line for $m>0$) or minimal (dashed red line for
    $-1<m<0$) values of the Lorentz
    number ${\cal L}$ (reached at $\mu=0$) obtained from
    Eq.\ (\ref{lormzeta}).
    Insets visualize the ${\cal T}(\varepsilon)$ function given by
    Eq.\ (\ref{temodm}) for $m=0$ and $m=1$, with contributions
    from the valence band $(p)$ and the conduction band $(n)$. 
  }
\end{figure}

We now elaborate possible effects, on the Lorentz number, of toy-models
of transmission-energy dependence
\begin{equation}
  \label{temodm}
  {\cal T}(\varepsilon) \propto |\varepsilon|^{m}, \qquad m>-1,
\end{equation}
where the proportionality coefficient is irrelevant due to the structure
of Eq.\ (\ref{lornano}). For some cases, integrals can be calculated
analytically, leading e.g.\ to ${\cal L}={\cal L}_0$ for $m=0$
(the constant transmission case), or to ${\cal L}={\cal L}(y)$ given by
Eq.\ (\ref{lorevials}) for $m=1$ (the linear transmission-energy dependence). 
Numerical results for selected values of $m=-0.5\,\dots{}\,2.5$ are displayed
in Fig.\ \ref{wftoys}. 

The violation of the WF law appears generically for $m\neq{}0$
away from the charge-neutrality point (i.e., for $\mu\neq{}0$). 

For $\mu=0$, the Lorentz number reaches a~global maximum (with
${\cal L}>{\cal L}_0$) if $m>0$, or a~global minimum (with
${\cal L}<{\cal L}_0$) if $-1<m<0$.
A~close-form expression can be derived for both the cases, 
namely
\begin{multline}
  \label{lormzeta}
  \frac{{\cal L}(\mu\!=\!0)}{\left(\dfrac{k_B}{e}\right)^2} \,=
  \frac{\displaystyle
  \int_0^\infty{}dx\dfrac{x^{m+2}}{\cosh^2\left(\frac{x}{2}\right)}}{ 
  \displaystyle
  \int_0^\infty{}dx\dfrac{x^m}{\cosh^2\left(\frac{x}{2}\right)}} = 
  \\
  \frac{2^{m+1}-1}{2^{m+1}-4}(m\!+\!1)(m\!+\!2)
  \frac{\zeta(m\!+\!2)}{\zeta(m)}, 
\end{multline}
and is visualized in Fig.\ \ref{lormax} \cite{mzet01foo}. 
It is clear that ${\cal T}(\varepsilon)$ models given by Eq.\
(\ref{temodm}) may lead to arbitrarily high ${\cal L}_{\rm max}$;
in particular, the value of $10\,{\cal L}_0$ is exceeded starting from
$m\approx{}4.1$.

Hence, for $m>1$, the model grasps the basic features of one-dimensional
Luttinger liquids, showing both the power-law transmission energy
dependence, with nonuniversal (interaction dependent) exponents,
and the significantly enhanced Lorentz numbers \cite{Wak11}. 

On the other hand, the suppression of ${\cal L}$ is
observed for $-1<m<0$, due to the integrable singularity at $\varepsilon=0$,
constituting an analogy with heavy fermion systems \cite{Tan07}. 

Both the above-mentioned scenarios were described theoretically
for quantum dot systems, which may be tuned from the suppression of
${\cal L}$ 
due to Breit-Wigner resonance, to the enhancement of ${\cal L}$
due to Fano resonance \cite{Sil12,Wan16,Kar20}.

\subsection{Gapped systems}
For a~sake of completeness, we show here how the energy (or transport)
gap may enhance the Lorentz number.
Instead of ${\cal T}(\varepsilon)$ given by Eq.\ (\ref{temodm}), we put
\begin{equation}
\label{temodgap}
  {\cal T}(\varepsilon) \propto
  \Theta\left(|\varepsilon|-{\textstyle\frac{1}{2}}\Delta\right)
  \left(|\varepsilon|-{\textstyle\frac{1}{2}\Delta}\right)^m,
  \quad m>-1, 
\end{equation}
where $\Theta(x)$ is the Heaviside step function.

For $\Delta\gg{}k_B{}T$ and $\Delta\gg{}|\mu|$, the integrals occuring
in Eq.\ (\ref{lornano}) can be approximated by elementary functions
(see {\em Supplementary Information} in Ref.\ \cite{Sus19})
and for the maximal ${\cal L}$ (reached for $\mu=0$) we have
\begin{equation}
  \label{lormaxgap}
  \frac{{\cal L}_{\rm max}}{\left(\dfrac{k_B}{e}\right)^2} \approx
  \left(\frac{\Delta}{2k_B{}T}+m+1\right)^2 + m+1. 
\end{equation}
This time, the result given in Eq.\ (\ref{lormaxgap}) can be simplified
in the $m\rightarrow{}-1$ limit, and takes the form of
${\cal L}_{\rm max}\approx{}(\Delta/2eT)^2$.
Physically, such a~limit is equivalent to the narrow band case, namely, 
${\cal T}(\varepsilon)\propto{}\delta(\varepsilon+{\textstyle\frac{1}{2}}\Delta)
+\delta(\varepsilon-{\textstyle\frac{1}{2}}\Delta)$,
with $\delta(x)$ being the Dirac delta function. 

An apparent feature of Eq.\ (\ref{lormaxgap}) is that ${\cal L}_{\rm max}$
shows an unbounded growth with a~gap (with the leading term being of
the order of $\sim{}\Delta^2$),
in agreement with the experimental results for semiconductors \cite{Gol56}. 
Similar behaviors can be expected for tunable-gap systems, such as
bilayer graphene or silicene, which are beyond the scope of this work.

\begin{figure}[!t]
  \includegraphics[width=\linewidth]{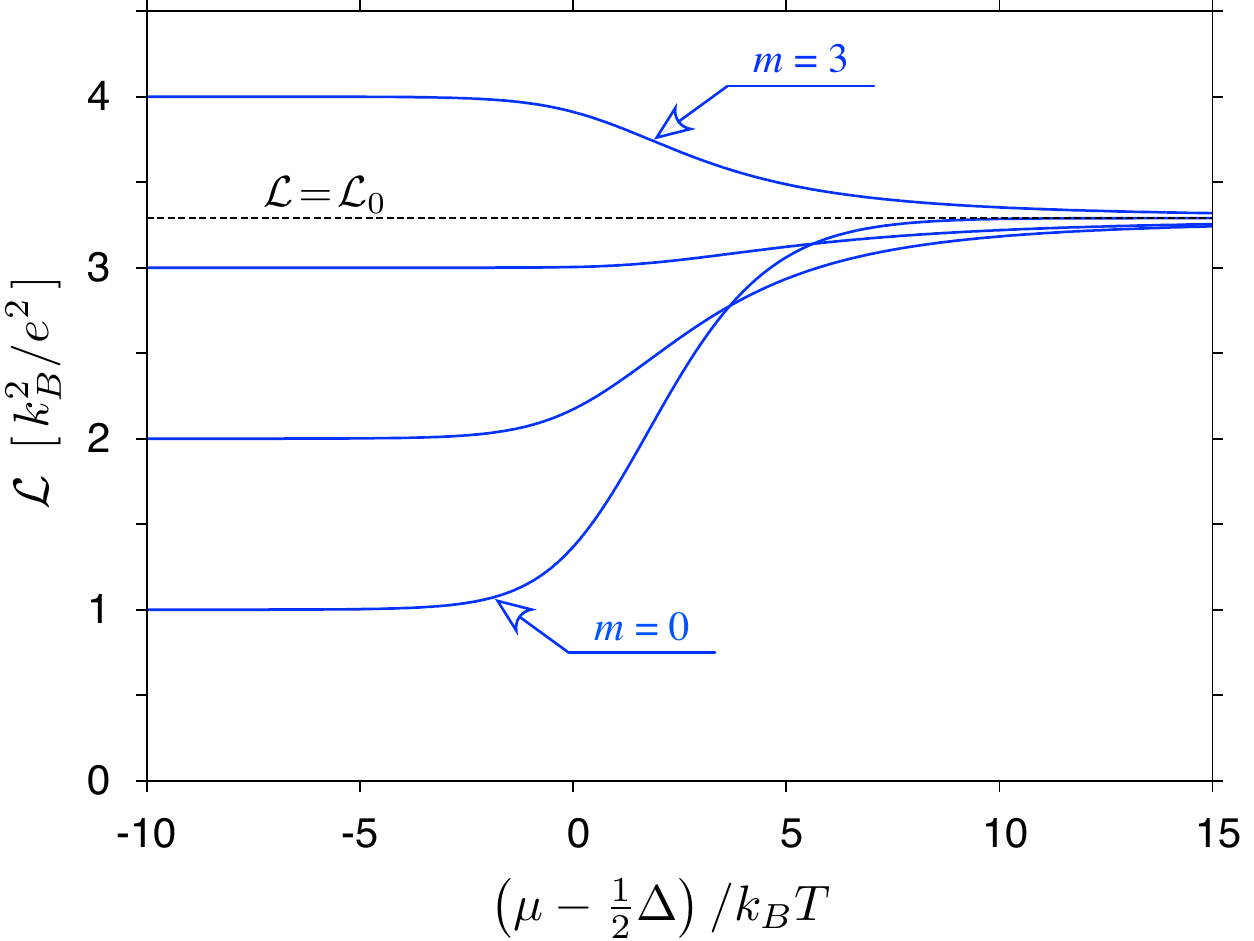}
  \caption{ \label{wftoun}
    The Lorentz number as a~function of chemical potential
    for the limit of an unipolar system, corresponding to
    ${\cal T}(\varepsilon)$ given by Eq.\ (\ref{temodgap})
    with $\Delta\gg{}k_B{}T$ and $\mu\approx{}\Delta/2$
    [see also Eq.\ (\ref{lorunitoy})].
    The exponent $m$ is varied from $0$ to $3$ with the steps
    of $1$ (solid lines).
    Dashed line marks the Wiedemann-Franz value (${\cal L}={\cal L}_0$). 
}
\end{figure}

A~different behavior appears near the band boundary, i.e., for 
$\mu\approx{}\Delta/2$ (or $\mu\approx{}-\Delta/2$).
Assuming $\Delta\gg{}k_B{}T$ again, we arrive to the limit of an unipolar
system, for which only the contribution from majority carries to integrals
$L_n$ (\ref{llndef}) matters.
In effect, the Lorentz number can be approximated
as
\begin{multline}
  \label{lorunitoy}
  {\cal L}(\mu\approx{\textstyle\frac{1}{2}\Delta\gg{}k_BT})
  \approx \\
  \left[\frac{{\cal J}_2(m;y)}{{\cal J}_0(m;y)}
  -\left(\frac{{\cal J}_1(m;y)}{{\cal J}_0(m;y)}\right)^2\right]
  \times\left(\frac{k_B}{e}\right)^2,
\end{multline} 
where $y=(\mu-\textstyle\frac{1}{2}\Delta)/k_B{}T$ and
\begin{equation}
  {\cal J}_n(m;y) = \int_{-y}^{\infty}dx\,x^n\frac{(x+y)^m}{\cosh^2(x/2)}. 
\end{equation}
Closed-form expressions for ${\cal J}_n(m;y)$ are not available; a~few
numerical examples for $m=0\dots{}3$ are presented in Fig.\ \ref{wftoun}.
Since now $L_1\propto{}{\cal J}_1\neq{}0$ (in contrast to the bipolar case
studied before), the Lorentz number is significantly reduced, and relatively
close to ${\cal L}_0$, which is approached for $y\gg{}1$. 

Asymptotic forms of  ${\cal J}_n(m;y)$ can be derived for $|y|\gg{}1$,
namely
\begin{align}
  {\cal J}_n(m;\,y\!\rightarrow{}\!-\!\infty) &\simeq{}
  4e^y\int_0^{\infty}{}dt(t-y)^nt^me^{-t}
  \nonumber \\
  &= 4e^y\sum_{k=0}^n{}{{n}\choose{k}}(-y)^k\Gamma(m\!+\!k\!+\!1), 
\end{align}
where $\Gamma(z)$ denotes the Euler gamma function, 
and
\begin{multline}
  {\cal J}_n(m;\,y\!\rightarrow{}\!+\!\infty) \simeq{}
  y^m\int_{-\infty}^{\infty}dx\frac{x^n}{\cosh^2(x/2)}
  \\
  = y^m\times
  \begin{cases}
    1, & \text{for }\ n=0, \\
    2\left(1\!-\!{2^{1-n}}\right)\Gamma(n\!+\!1)\zeta(n),
    & \text{for }\ n\geqslant{}1. \\
  \end{cases}
\end{multline}
Substituting the above into Eq.\ (\ref{lorunitoy}), we obtain
\begin{equation}
  {\cal L}\rightarrow{}(m+1)\left(\frac{k_B}{e}\right)^2\ \
  \text{for }\ \ y\rightarrow{}-\infty, 
\end{equation}
or
\begin{equation}
  {\cal L}\rightarrow\frac{\pi^2}{3}\left(\frac{k_B}{e}\right)^2
  ={\cal L}_0,\ \ 
  \text{for }\ y\rightarrow{}\infty.  
\end{equation}
Both limits are closely approached by the numerical data
in Fig.\ \ref{wftoun} for $|y|\gtrsim{}5$. 
In all the cases considered, the values of ${\cal L}$ are now much lower
than the corresponding ${\cal L}_{\rm max}$ for a~gapless model with the same
$m$ (see Fig.\ \ref{lormax}).

Therefore, it becomes clear from analyzing simplified models
of ${\cal T}(\varepsilon)$ that a~bipolar nature of the system, next
to the monotonically-increasing transmission (the $m>0$ case) are
essential when one looks for a~significant enhancement of
the Lorentz number ${\cal L}$ (compared to ${\cal L}_0$).

Both these conditions are satisfied for graphene.

%%\section{Exactly solvable mesoscopic systems in graphene}
\section{Exactly solvable mesoscopic systems} 
\label{mesosys}

\subsection{Transmission-energy dependence}
The exact transmission-energy dependence ${\cal T}(\varepsilon)$
can be given for two special device geometries in graphene:
a~rectangular sample attached to heavily-doped graphene leads
\cite{Kat06,Two06,Pra07} and for the Corbino disk \cite{Ryc09,Ryc10}.
Although these systems posses peculiar symmetries, allowing one to solve
the scattering problem employing analytical mode-matching method
(in particular, the mode mixing does not occur), both the solutions were
proven to be robust against various symmetry-breaking perturbations
\cite{Bar07,Lew08,Sus20a}. 
More importantly, 
several features of the results have been confirmed in the experiments 
\cite{Mia07,Dan08,Kum18,Zen19} showing that even such idealized systems
provide valuable insights into the quantum transport phenomena involving
Dirac fermions in graphene. 

For a~rectangle of width $W$ and length $L$, the transmission can be
written as \cite{Two06,Ryc09}
\begin{equation}
  {\cal T}(\varepsilon) = \sum_{n=0}^{\infty}T_n, 
\end{equation}
where the transmission probability for $n$-th normal mode is given by
\begin{equation}
  \label{ttnrect}
  T_n = \left[
    1+\left(\dfrac{q_n}{k_n}\right)^2\sin^2\left(k_n{}L\right)
  \right]^{-1}, 
\end{equation}
with $q_n=\pi(n+\frac{1}{2})/W$ the quantized transverse wavevector
(the constant $\frac{1}{2}$ corresponds to infinite-mass confinement;
for other boundary conditions, see Ref.\ \cite{Two06}), 
\begin{equation}
  \label{kncases}
  k_n = \begin{cases}
  \sqrt{k^2-q_n^2}, & \text{for }\  k\geqslant{}q_n, \\
  i\sqrt{q_n^2-k^2}, & \text{for }\  k<q_n, \\
  \end{cases}
\end{equation}
and $k=|\varepsilon|/(\hbar{}v_F)$. The two cases in Eq.\ (\ref{kncases})
refer to the contributions from propagating waves ($k\geqslant{}q_n$,
so-called {\em open channels}) and evanescent waves ($k<q_n$). 

For the Corbino disk, with its inner ($R_1$) and outer ($R_2$) radii, 
we have \cite{Ryc09} 
\begin{equation}
  {\cal T}(\varepsilon) = \sum_{j=\pm{}1/2,\pm{}3/2,\dots}T_j, 
\end{equation}
where $j$ the the half-odd integer angular momentum quantum number,
with a~corresponding transmission probability
\begin{equation}
\label{tjphi}
  T_{j} = \frac{16}{\pi^2{}k^2{}R_1{}R_2}\,
  \frac{1}{\left[\mathfrak{D}_{j}^{(+)}\right]^2
    + \left[\mathfrak{D}_{j}^{(-)}\right]^2},
\end{equation}
where $k$ is same as in Eq.\ (\ref{kncases}), and  
\begin{multline}
\label{ddnupm}
  \mathfrak{D}_{j}^{(\pm)} = \mbox{Im}\left[
    H_{j-1/2}^{(1)}(kR_1)H_{j\mp{}1/2}^{(2)}(kR_2)\right. \\
    \pm \left.H_{j+1/2}^{(1)}(kR_1)H_{j\pm{}1/2}^{(2)}(kR_2)
    \right], 
\end{multline}
with $H_{\nu}^{(1,2)}(\rho)$ the Hankel function of the (first, second) kind.

\subsection{The conductivity}
A~measurable quantity that provides a~direct insight into the
${\cal T}(\varepsilon)$ function is zero-temperature conductivity
\begin{equation}
  \label{sigzero}
  \sigma(\varepsilon) = g_0\Omega_{X} {\cal T}(\varepsilon), 
\end{equation}
with the conductance quantum $g_0=4e^2/h$ and a~shape-dependent factor
\begin{equation}
  \Omega_X =
  \begin{cases}
  {L}/{W}, & \text{for rectangle}, \\
  \dfrac{1}{2\pi}\ln\left(R_2/R_1\right), & \text{for disk}. \\
  \end{cases}
\end{equation}
For $T>0$, Eq.\ (\ref{sigzero}) needs to be replaced by 
$\sigma(\mu) = e^2\,\Omega_X{}L_0$,
where $L_0$ is given by Eq.\ (\ref{llndef}) with $n=0$.

\begin{figure}[!t]
  \includegraphics[width=\linewidth]{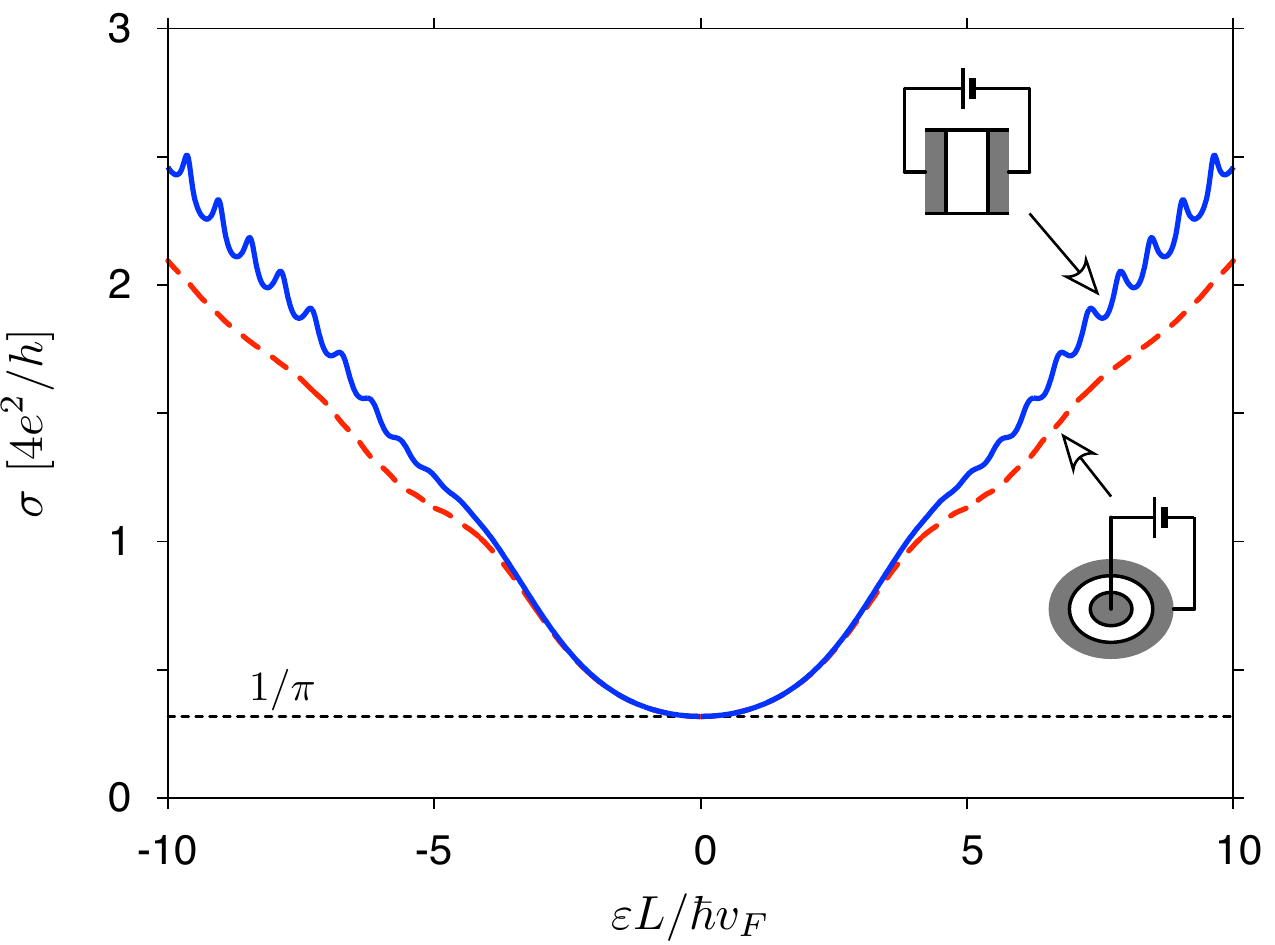}
  \caption{ \label{sigland}
    Zero-temperature conductivity as a~function of the Fermi energy
    for a~rectangular sample with width-to-length ratio $W/L=5$
    (solid blue line) and the Corbino disk with radii ratio $R_2/R_1=2$
    (dashed red line). Both system are shown schematically.
    Dashed black line marks the universal conductivity
    $\sigma_0=(4/\pi)\,e^2/h$. 
  }
\end{figure}

Numerical results, for $T=0$, are presented in Fig.\ \ref{sigland}.
The data for both systems, displayed versus a~dimensionless quantity
$\varepsilon{}L/\hbar{}v_F$ (with $L\equiv{}R_2-R_1$ for a~disk) 
closely follow each other up to $|\varepsilon{}|L/\hbar{}v_F\approx{}3$.
For larger values of $|\varepsilon|$, the results become shape-dependent 
and can be approximated, for $|\varepsilon|\gg\hbar{}v_F/L$, as
\begin{equation}
  \label{sigball}
  \sigma(\varepsilon) \approx{} g_0\Omega_X
  N_{\rm open}(\varepsilon)\langle{T}\rangle_{\rm open},
\end{equation}
where the number of open channels
\begin{equation}
  \label{nopenflo}
   N_{\rm open}(\varepsilon) =
   \begin{cases}
     \lfloor{kW/\pi}\rfloor, & \text{for rectangle}, \\
     2\lfloor{kR_1}\rfloor, & \text{for disk}, \\
   \end{cases}
\end{equation}
with $\lfloor{}x\rfloor$ being the floor function of $x$, and the
average transmission per open channel $\langle{T}\rangle_{\rm open}\approx\pi/4<1$
(for the derivation, see {\it Appendix~A}).
Remarkably, numerical values of $\sigma(\varepsilon)$ for a~rectangle with
$W/L=5$ [solid blue line in Fig.\ \ref{sigland}] match the approximation
given by Eq.\ (\ref{sigball}) with a~few percent accuracy for
$|\varepsilon|\gtrsim{}\,5\hbar{}v_F/L$, whereas for a~disk with $R_2/R_1=2$
[dashed red line] a~systematic offset of $\approx{}(1/\pi)g_0$ occurs,
signaling an emphasized role of evanescent waves in the Corbino geometry.
This observation coincides with a~total lack of Fabry-Perrot oscillations
in the Corbino case.

\begin{figure}[!t]
  \includegraphics[width=\linewidth]{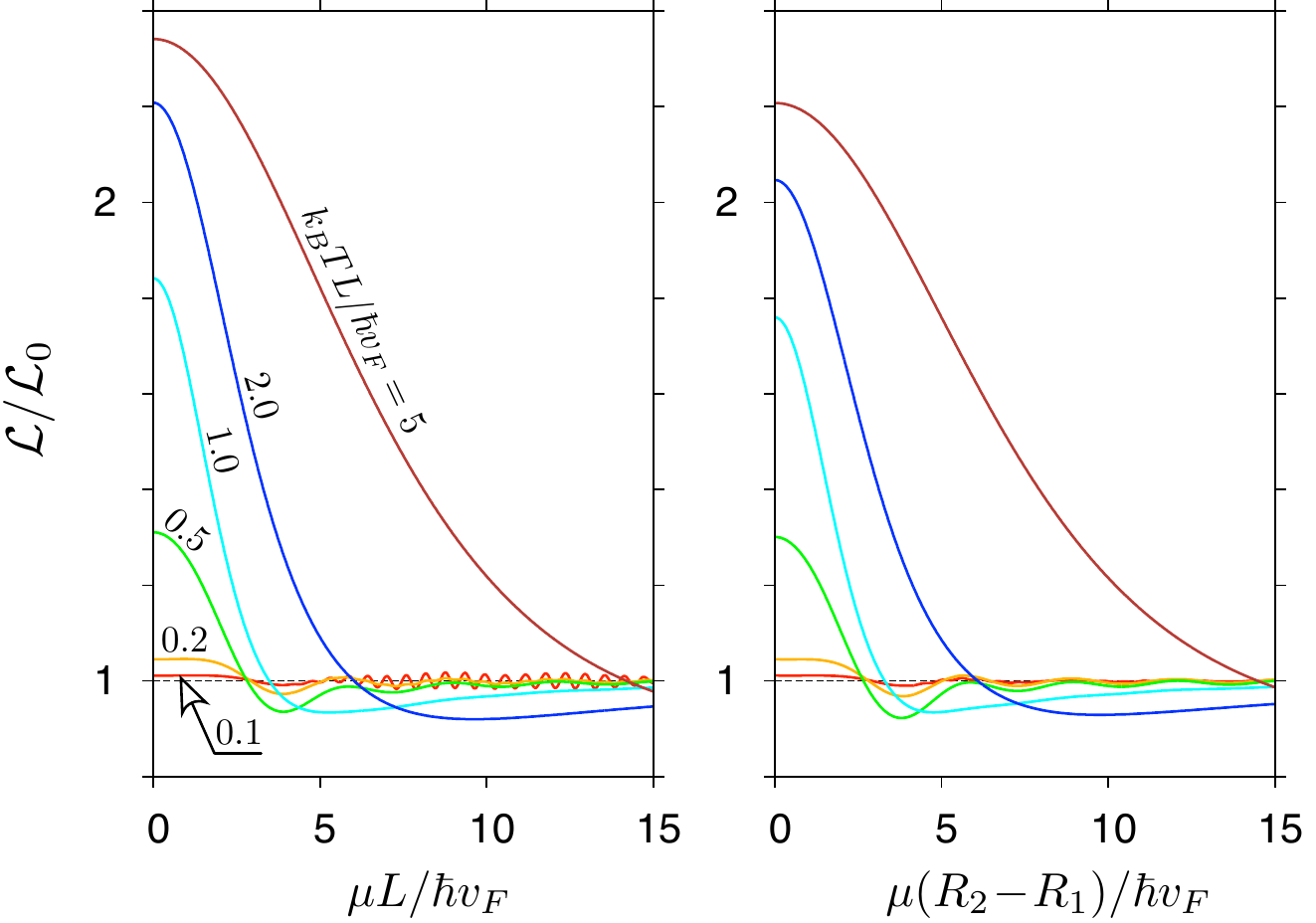}
  \caption{ \label{wfland}
    The Lorentz number for a~rectangular sample (left) and the Corbino disk
    (right) displayed as a~function of the chemical potential.
    The temperature, specified in the units of $\hbar{}v_F/(k_BL)\approx{}
    6.67\,$K$\,\cdot\,\mu$m$\,\times{}\,L^{-1}$, is varied between
    the lines and same in both panels.
    Remaining parameters are same as in Fig.\ \ref{sigland}. 
  }
\end{figure}

\begin{figure}[!t]
  \includegraphics[width=\linewidth]{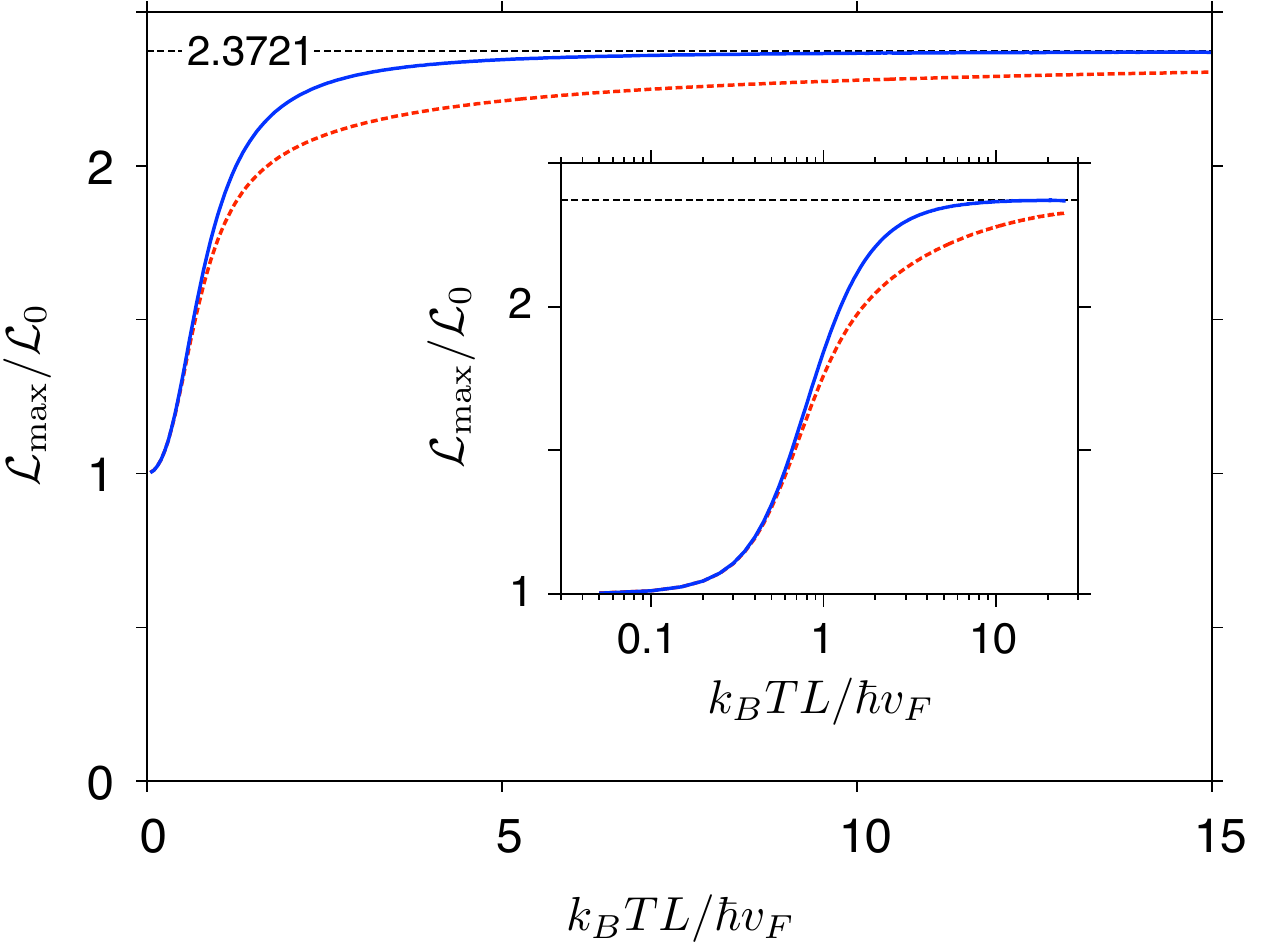}
  \caption{ \label{wfland0kt}
    Maximal Lorentz number (corresponding to $\mu=0$) for same systems
    as in Fig.\ \ref{sigland} versus temperature. 
    Inset shows the data replotted from main panel with the abscissa
    scaled logarithmically. Dashed horizontal line marks the prediction
    given in Eq.\ (\ref{lor0lin}). 
  }
\end{figure}

\subsection{The Lorentz number}
The exact transmission-energy functions ${\cal T}(\varepsilon)$, discussed
above, are now substituted  to Eq.\ (\ref{lornano}) for the Lorentz number.
Calculating the relevant integrals numerically, 
we obtain the results presented in Figs.\ \ref{wfland} and \ref{wfland0kt}.

Close to the charge-neutrality point, i.e., for $|\mu|\lesssim{}
$max$(\hbar{}v_FL^{-1},k_B{}T)$, both systems show a~gradual crossover
(with increasing $T$) from the Wiedemann-Franz regime, with a~flat
${\cal L}\approx{}{\cal L}_0$, to the linear-transmission regime characterized
by ${\cal L}(\mu)$ close to the predicted by Eq.\ (\ref{lorevials})
[see Fig.\ \ref{wfland}]. 
For higher $\mu$, some aperiodic oscillations of ${\cal L}(\mu)$  are
visible if $k_BT\lesssim{}\hbar{}v_F/L$, being particularly well pronounced
for a~rectangular sample.
For higher temperatures, the oscilltions are smeared out, leaving only
one shallow minimum near $|\mu|/k_B{}T\approx{}4-5$, in agreement with
Eq.\ (\ref{lorevials}). 

Maximal values of ${\cal L}$ for the two systems (reached at $\mu=0$)
are displayed, as functions of temperature, in Fig.\ \ref{wfland0kt}. 
It is clear that a~crossover between low and high temperature regimes
takes place near $k_B{}T\sim{}\hbar{}v_F/L$ (corresponding to
$\approx{}6.67\,$K for $L=1\,\mu$m):
For lower temperatures
(and near $\mu=0$), thermally-excited carriers appear in the area where
${\cal T}(\varepsilon)\approx{}$const (leading to ${\cal L}\approx{}
{\cal L}_0$), whereas for significantly higher temperatures, the detailed
behavior of ${\cal T}(\varepsilon)$ near $\varepsilon=0$ becomes irrelevant, 
and the linear-transmission approximation ($\,{\cal T}(\varepsilon)
\propto|\varepsilon|\,$) applies.
Remarkably, the convergence to the value given in Eq.\ (\ref{lor0lin})
is much slower (yet clearly visible) in the Corbino disk case, due to
a~higher (compared to a~rectangular sample) contribution from evanescent
waves to the transmission away from the charge-neutrality point.

\section{Conclusions}
\label{conclu}

We have calculated the Lorentz number (${\cal L}=\kappa_{\rm el}/\sigma{}T$)
for noninteracting massless Dirac fermions following two different
analytic approaches:
{\em first}, adapting the handbook derivation of the Wiedemann-Franz (WF) law,
starting from the relation between thermal conductivity and heat
capacity obtained within the kinetic theory of gases, 
and {\em second}, involving the Landauer-B\"{u}ttiker formalism
and postulating simple model of transmission-energy dependence,
${\cal T}(\varepsilon)
\propto{}|\varepsilon|$.
In both approaches, the information about conical dispersion relation
is utilized, but the universal value of electrical conductivity, 
$\sigma\sim{}e^2/h$ at $\varepsilon=0$,
is referred only in the first approach.
Nevertheless, the results are numerically close, indicating the violation
of the WF law with maximal Lorentz numbers ${\cal L}_{\rm max}/{\cal L}_0
\approx{}2.77$ and $2.37$ (respectively) and ${\cal L}\rightarrow
{\cal L}_0=(\pi^2/3)\,k_B^2/e^2$ for high doppings ($|\varepsilon|\gg{}k_B{}T$).
This observation suggests that violation of the WF law, with
${\cal L}_{\rm max}/{\cal L}_0\approx{}2-3$ should appear generically
in weakly-doped systems with approximately conical dispersion relation, including
multilayers and hybrid structures, even when low-energy details of the band
structure alter the conductivity. 

Moreover, a~generalized model of power law transmission-energy dependence, 
${\cal T}(\varepsilon)\propto{}|\varepsilon|^m$ (with $m>-1$), 
is investigated in order to  address the question whether the enhancement
of ${\cal L}$ is due to the bipolar band structure or due to the conical
dispersion. Since ${\cal L}>{\cal L}_0$ shows up for any $m>0$, and
the maximal value grows monotonically with $m$, we conclude that
the dispersion relation has a~quantitative impact on the effect.
On the other hand, analogous discussion of gapped systems, with the chemical
potential close to the center of the gap (the bipolar case) or to
the bottom of the conduction band (the unipolar case) proves that the bipolar
band structure is also important (no enhancement of ${\cal L}$ 
is observed in the unipolar case up to $m\approx{}2$). 

Finally, the Lorentz numbers, for different dopings and temperatures,
are elaborated numerically from exact solutions available
for the rectangular sample and the Corbino (edge-free) disk in graphene, both
connected to heavily-doped graphene leads. 
The results show that ${\cal L}$, as a~function of the chemical potential
$\mu$, gradually evolves (with growing $T$) as expected for
a~model transmission energy dependence,
${\cal T}(\varepsilon)\propto{}|\varepsilon|^m$,
with the exponent varying from $m=0$ to $m=1$. 
The upper bound is approached faster for the rectangular sample case,
but in both cases ${\cal L}/{\cal L}_0>2$ is predicted to appear for
$T\gtrsim{}13\,\text{K}\cdot{}\mu\text{m}\times{}L^{-1}$ with $L$ the sample length. 

Our results complement earlier theoretical study on the topic \cite{Yos15}
by including the finite size-effects and the interplay between propagating
and evanescent waves, leading to the results dependent, albeit weakly, 
on the sample geometry.

\section*{Acknowledgments}
The work was supported by the National Science Centre of Poland (NCN)
via Grant No.\ 2014/14/E/ST3/00256.
Discussions with Manohar Kumar are appreciated.

\appendix

\section{Average transmission per open channel and the enhanced shot noise
  away from the Dirac point}

\begin{figure}[t]
  \includegraphics[width=\linewidth]{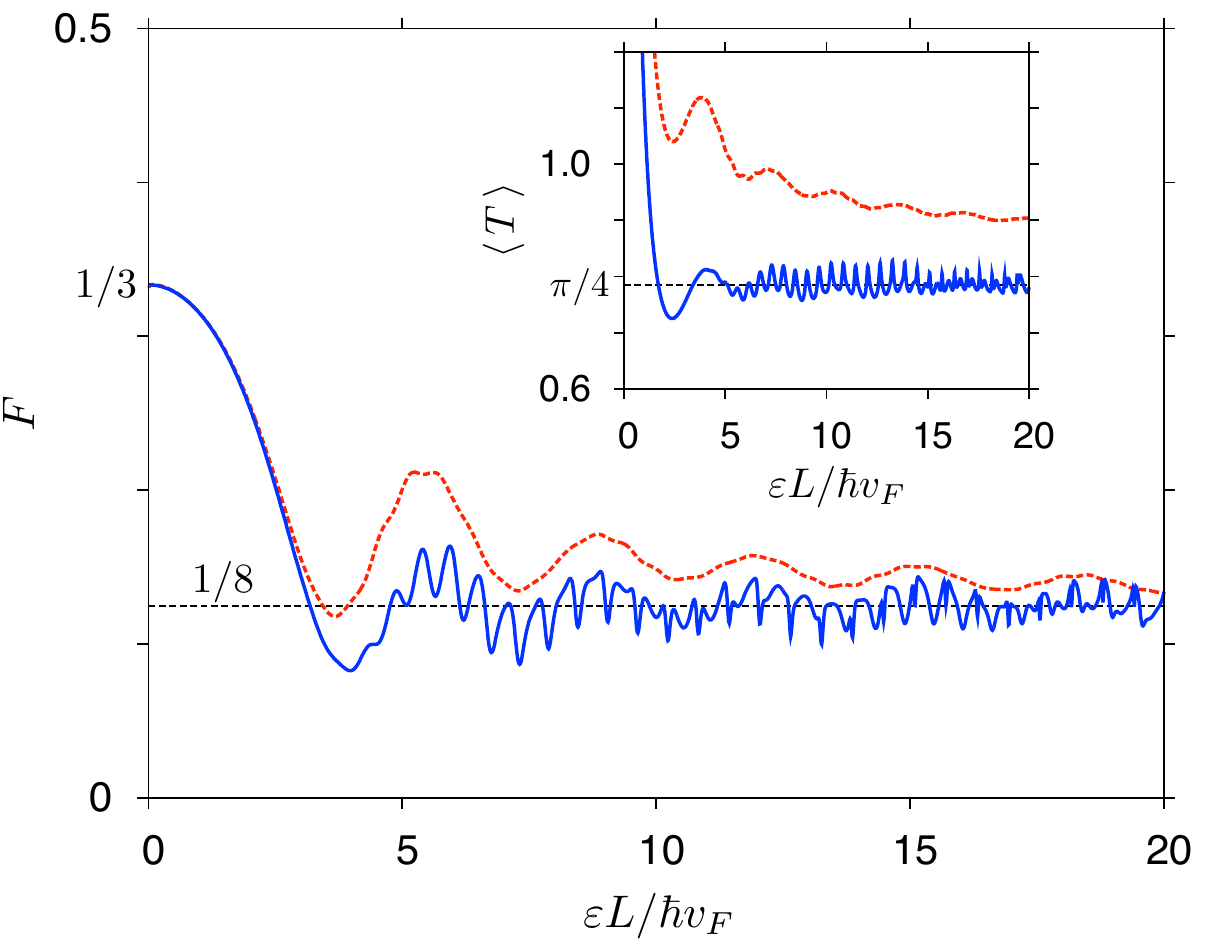}
  \caption{ \label{fanoland}
    Fermi energy dependence of the Fano factor (main panel) and
    the average transmission per channel (inset), defined
    in Eq.\ (\ref{avettdef}), for same systems as in Fig.\ \ref{sigland}. 
  }
\end{figure}

In this Appendix, we explain why the average transmission per open channel,
occurring in Eq.\ (\ref{sigball}) in the main text, is 
$\langle{T}\rangle_{\rm open}\approx\pi/4$ instead of $1$ (being the value
expected for typical ballistic systems).
Implications for the shot-noise power are also briefly discussed. 

A~closer look at Eq.\ (\ref{ttnrect}) for the transmission probability
allows us to find out, for high energy,  that $k_nL\gg{}1$ (typically)
whereas $q_n$ and $k_n$, for open channels, are bounded by $k$.
Therefore, the average
transmission can be approximated by replacing the argument of sine
by a~random phase $0\leqslant\varphi<\pi$, and taking averages over
$\varphi$ and $n$ independently, 
\begin{align}
  \label{ttpi4}
  \langle{T}\rangle_{\rm open} &\approx
  \frac{1}{\pi}\int_0^{\pi}d\varphi{}\int_0^1{}dx
  \frac{1}{1+\frac{x^2}{1-x^2}\sin^2{}\varphi}
  \nonumber \\
  &= \int_0^1{}dx\sqrt{1-x^2} = \frac{\pi}{4}, 
\end{align}
where we have further introduced a~continuous parametrization
$x=q_n/k$, $\sqrt{1-x^2}=k_n/k$. 
In analogous way we obtain
\begin{align}
  \langle{T^2}\rangle_{\rm open} &\approx
  \frac{1}{\pi}\int_0^{\pi}d\varphi{}\int_0^1{}dx
  \frac{1}{\left(1+\frac{x^2}{1-x^2}\sin^2{}\varphi\right)^2}
  \nonumber \\
  &= \frac{7\pi}{32}. 
\end{align}

The Fano factor \cite{Two06}, quantifying the shot-noise power,
can now be approximated, for $kL\gg{}1$, as
\begin{equation}
  \label{fanosum}
  F = \frac{\sum_nT_n(1-T_n)}{\sum_nT_n}
  \approx{}1-\frac{\langle{T^2}\rangle_{\rm open}}{\langle{T}\rangle_{\rm open}}
  =\frac{1}{8}. 
\end{equation}
The last value in Eq.\ (\ref{fanosum}) indicates that shot-noise power
in highly-doped graphene is
noticeably enhanced comparing to standard ballistic systems,
which are characterized by $F\approx{}0$ (as $T_n=0$ or $1$
for all modes).

Exact results, obtained from first equality in Eq.\ (\ref{fanosum}),
taking both propagating and evanescent modes into account, are presented
in Fig.\ \ref{fanoland}.
The average transmission, displayed in the inset, is defined as
\begin{equation}
  \label{avettdef}
  \langle{}T\rangle = 
  \frac{{\cal T}(\varepsilon)}{\widetilde{N}_{\rm open}(\varepsilon)}, 
\end{equation}
where $\widetilde{N}_{\rm open}(\varepsilon)$ is calculated from
Eq.\ (\ref{nopenflo}) in the main text in which we have omitted the
floor function (in general, $\widetilde{N}_{\rm open}(\varepsilon)
\geqslant{}{N}_{\rm open}(\varepsilon)$).

It is clear from Fig.\ \ref{fanoland} that a~stronger role of evanescent
modes for the Corbino case results in elevated $F$ and $\langle{}T\rangle$
(comparing to a~rectangle), but a~gradual convergence with growing
$\varepsilon$ to the values given by last equalities in Eqs.\ (\ref{fanosum})
and (\ref{ttpi4}) (respectively) is also visible. 

It is worth to notice that experimental values of $F\approx{}0.15$
for highly-doped graphene samples \cite{Dan08} are close, but slightly
elevated in comparison to $F\approx{}1/8$ in Eq.\ (\ref{fanosum}). 
This can be attributed to the tunneling assisted by charged impurities, 
or other defects, which may amplify the role of evanescent modes also
for rectangular samples.

%%%%%%%%%%%%%%%%%%%%%%%%%%%%%%%%%%%%%%%%%%%%%%%%%%%%%%%%%%%%%%%%%%%%%%%%%%%%%%
%%%%%%%%%%%%%%%%%%%%%%%%%%%%%%%%%%%%%%%%%%%%%%%%%%%%%%%%%%%%%%%%%%%%%%%%%%%%%%


\begin{thebibliography}{}
\bibitem{strafoo}
Strictly speaking, the value of $v_F$ may also be modified (by up to
$20-30\,\%$) by applying strain. However, controlling $v_F$ is much 
more difficult than controlling $n$ via the gate voltage.
\bibitem{mzet01foo}
The analytic continuation for $m\rightarrow{}0$ and $m\rightarrow{}1$
is assumed in the righmost expression in Eq.\ (\ref{lormzeta}).
\end{thebibliography}
\end{document}